\documentclass[sigconf,nonacm,screen]{acmart}

\usepackage{csquotes}
\usepackage{comment}
\usepackage{orcidlink}
\usepackage{mdframed}
\usepackage{cleveref}
\usepackage{booktabs}

\usepackage[inline]{enumitem}

\newcommand{\researchquestion}[2]{
\noindent
\fbox{\begin{minipage}{\dimexpr\linewidth-2\fboxsep-2\fboxrule\relax}
\textbf{#1:} \textit{\enquote{#2}}
\end{minipage}}
\vspace{2mm}
}

\AtBeginDocument{%
  \providecommand\BibTeX{{%
    \normalfont B\kern-0.5em{\scshape i\kern-0.25em b}\kern-0.8em\TeX}}}

\setcopyright{acmcopyright}
\copyrightyear{2024}
\acmYear{2024}
\acmDOI{XXXXXXX.XXXXXXX}

\acmConference[ICSE 2024]{46\textsuperscript{th} International Conference on Software Engineering}{April 14--20,
  2024}{Lisbon, Portugal}
\acmBooktitle{ICSE '24: 46\textsuperscript{th} International Conference on Software Engineering,
 April 14--20, 2024, Lisbon, Portugal} 
\acmPrice{15.00}
\acmISBN{978-1-4503-XXXX-X/18/06}

\begin{document}

\title{From Requirements to Architecture: An AI-Based Journey to Semi-Automatically Generate Software Architectures}

\author{Tobias Eisenreich\orcidlink{0009-0004-7168-251X}}
\email{eisenreich@iste.uni-stuttgart.de}
\orcid{0009-0004-7168-251X}
\affiliation{%
  \institution{Institute of Software Engineering}
  \city{Stuttgart}
  \postcode{70569}
  \country{Germany}
}

\author{Sandro Speth\orcidlink{0000-0002-9790-3702}}
\email{sandro.speth@iste.uni-stuttgart.de}
\orcid{0000-0002-9790-3702}
\affiliation{%
  \institution{Institute of Software Engineering}
  \city{Stuttgart}
  \postcode{70569}
  \country{Germany}
}

\author{Stefan Wagner\orcidlink{0000-0002-5256-8429}}
\email{stefan.wagner@iste.uni-stuttgart.de}
\orcid{0000-0002-5256-8429}
\affiliation{%
  \institution{Institute of Software Engineering}
  \city{Stuttgart}
  \postcode{70569}
  \country{Germany}
}

\renewcommand{\shortauthors}{Eisenreich et al.}

\begin{abstract}
Designing domain models and software architectures represents a significant challenge in software development, as the resulting architectures play a vital role in fulfilling the system's quality of service.
Due to time pressure, architects often model only one architecture based on their known limited domain understanding, patterns, and experience instead of thoroughly analyzing the domain and evaluating multiple candidates, selecting the best fitting.
Existing approaches try to generate domain models based on requirements, but still require time-consuming manual effort to achieve good results.
Therefore, in this vision paper, we propose a method to generate software architecture candidates semi-automatically based on requirements using artificial intelligence techniques.
We further envision an automatic evaluation and trade-off analysis of the generated architecture candidates using, e.g., the architecture trade-off analysis method combined with large language models and quantitative analyses.
To evaluate this approach, we aim to analyze the quality of the generated architecture models and the efficiency and effectiveness of our proposed process by conducting qualitative studies.
\end{abstract}



\maketitle

\section{Introduction}
Software architecture plays a vital role in the quality of every software system.
However, deriving a software architecture model fitting the functional and non-functional requirements is challenging, especially for complex and unknown domains, as it requires domain knowledge and understanding.
Especially in modern methods like Domain-Driven Design~\cite{evans2004domain}, building an elaborate domain model constitutes the basis for deriving the architecture.
In addition, architecture creation is very time-consuming and often based on incomplete or contradictory requirements documents.
New software architects especially lack experience and technical knowledge, but more experienced ones may focus more on concepts and technologies they are used to.

Hence, architects often design one architecture instead of multiple architecture candidates, which are evaluated against the requirements and other factors.
Thus, mistakes during the design phase of the architecture can show up later, imposing high maintenance costs.
Therefore, we require means to support software architects in their architectural design decisions and evaluation to reduce the time required and make the resulting architectures more reproducible through semi-automation.

In related work, Kof~\cite{kof2005text} tries to generate domain models as a basis for the architecture from requirements.
However, his process still requires a lot of time-consuming manual work to generate a fitting domain model.
Due to the rapidly growing capabilities of modern artificial intelligence (AI) techniques, e.g., natural language processing (NLP), such as large language models, we believe that we can reduce the manual effort substantially by applying these techniques in a semi-automated way.

We propose a method, i.e., a process and tool, to generate architectures semi-automatically based on requirements using modern AI techniques.
In our method, we first generate domain models and use-case scenarios using large language models.
Then, we derive multiple software architecture candidates with their architectural decision decisions and evaluate them to make informed decisions about which candidate to take. 
We include manual iterations to improve the overall outcome.
We believe that this process can potentially improve the quality of architecture and the creation time in most projects.

It may seem counter-intuitive to restrict the AI to the standard procedure.
However, the standard procedure has the advantage that it can easily be applied semi-automatically, and an experienced architect can nudge the LLM in the right direction.
Our exploratory analysis in \cref{sec:exploratory-analysis} suggests that a fully automated solution might not perform well with current state-of-the-art LLMs.
Furthermore, no data is available to train LLMs for this task, so we cannot easily improve upon the standard LLMs.
Our approach has the advantage of relying on general-purpose LLMs without special training.

To evaluate our approach, we aim to answer the following research questions:

\researchquestion{RQ1}{%
    Can state-of-the-art NLP technology generate reproducible, correct, and elaborate domain models and use case scenarios based on requirements in natural language?%
}\\
\researchquestion{RQ2}{%
    Can state-of-the-art AI technology generate software architectures based on a domain model, use case scenarios, and requirements that can appropriately fulfill these requirements?%
}\\
\researchquestion{RQ3}{%
    Can quantitative and qualitative software architecture evaluations and trade-off analyses be automated through the use of AI?%
}\\
\researchquestion{RQ4}{%
    Does a method for semi-automatic architecture generation improve the architecture's quality while reducing the time required?%
}

The remainder of this paper is structured as follows:
First, we outline the related work in \Cref{sec:related-work}.
In \Cref{sec:vision}, we describe our proposed approach.
In \Cref{sec:exploratory-analysis}, we describe an exploratory analysis we have conducted.
Finally, we explain our planned evaluation in \Cref{sec:evaluation} and conclude in \Cref{sec:conclusion}.

\section{Related Work}\label{sec:related-work}

Souza et al.~\cite{souza2019deriving} reviewed the state-of-the-art of architecture derivation from requirement specifications.
They found 39 relevant studies, some of which included methods with tooling support, but all of them relied heavily on the experience and knowledge of the architect.
Kof~\cite{kof2005text} took an approach to more automated support.
He developed a semi-automatic method to create domain models from requirements documents.
Despite the semi-automatic approach, his method took extensive manual effort, especially for preparing the requirements documents.

Researchers have touched on automated architecture evaluation before:
Bashroush et al.~\cite{bashroush2004towards} manage to apply ATAM~\cite{kazman2000atam} with partial automation.
This accelerates the process of the evaluation.
They use a formal Architecture Description Language for this task.
Kr??ger et al.~\cite{kruger2006automating} also try to speed up the architecture evaluation, but go down a different route:
They developed a method to transition scenarios into AspectJ prototypes.
With these prototypes, multiple architectures can be manually evaluated and compared during runtime.
Scheerer et al.~\cite{scheerer2017automatic} use PerOpteryx, a Palladio~\cite{becker2009palladio} plugin, to run automated quantitative analysis on architectures.
They evaluate architecture design decisions by analyzing their performance impact for predefined scenarios.
Numerous works evaluate architectures based on metrics;
Coulin et al.~\cite{coulin2019software} provide an overview of the state of the art.
Mo et al.~\cite{mo2018experiences} successfully find architectural flaws in industrial projects with fully automated architecture metrics.
Other works try to optimize architectures:
Aleti et al.~\cite{aleti2012software} conducted a systematic literature review of 188 papers that evaluate architecture optimization methods.
To the best of our knowledge, there is no automatic and comprehensive architecture evaluation that takes the requirements fully into account.

\begin{figure*}[tbp]
    \centering
    \includegraphics[width=\textwidth]{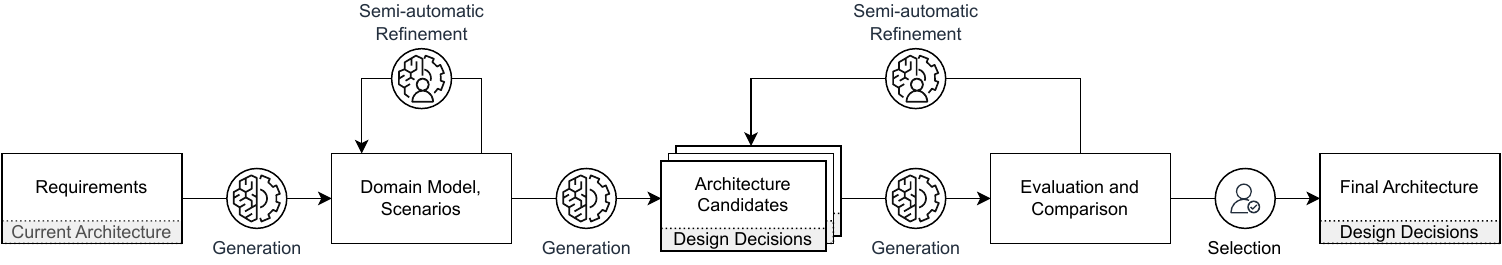}
    \caption{Semi-automatic architecture creation process.}
    \label{fig:process}
\end{figure*}

\section{Vision for Semi-Automatic Architecture Generation}\label{sec:vision}

In the following, we propose our vision for a process and tool framework to generate software architectures semi-automatically based on requirements. 
Our process is envisioned to be iterative throughout a software development life-cycle.
\Cref{fig:process} depicts this process and the relevant artifacts.
The process consists of six steps: (1)~automatically generating a domain model and use-case scenarios based on textual requirements, (2)~manually refining the domain model and scenarios, (3)~automatically deriving multiple software architecture candidates and architectural decisions leading to them using the domain model, scenarios, and non-functional requirements, (4)~automatically evaluating and comparing the candidates, (5)~manually refining the candidates, and (6)~manually selecting the best fitting candidate.
In the following, we will describe each step in more detail.

\subsection{Generating Domain Model and Use-Case Scenarios}
In the first step, we want to generate a domain model and use-case scenarios out of functional and non-functional requirements.
While we only consider textual requirements in the first iteration, in later ones, we also include the current architecture and architecture decisions leading to it.
This will allow us to apply this process to incremental architecture.

Using a large language model (LLM), the software architect processes the requirements to create an initial domain model and use-case scenarios.
We deem LLMs suitable for such a task as they provide decent knowledge in plenty of domains and, thus, are universally usable in contrast to more domain-specific AI models.
A first exploratory analysis with LLaMA showed promising results (c.f. \Cref{sec:exploratory-analysis}).
To be able to fine-tune the LLM if required, we plan to use an open-source LLM for this task.
Currently, we are considering LLaMA, Falcon, and Yi, but we will evaluate the state-of-the-art models for their suitability as part of our research, potentially including other models that have not been published yet.
Furthermore, we will compare these LLMs' results with those obtained from the latest GPT version.

The generated domain model will omit details that are irrelevant to the architecture generation, e.g., an entity's attributes.
The domain model's primary focus lies in the relations of the domain concepts to each other.
With these relations, we can cut the domain into bounded contexts.
We can then cut the architecture into its components with these bounded contexts.
The use case scenarios will complement the domain model with the description of the system's functional behavior.

After generating the domain model and scenarios, the architect can inspect and refine them using additional prompts.
We assume that the domain model and the scenarios can be refined using the LLM, so the architect does not need to do any work directly on the models but instead instructs the tooling.

\subsection{Generating Architecture Candidates and Architecture Decision Records}
In the third step, we want to generate multiple architecture candidates based on the requirements, the domain model, and the use-case scenarios.
These architecture candidates model the different design decisions that are viable to take when considering the requirements, the domain model, and the use-case scenarios.

We will consider various models for the architecture:
First, there is the classic \enquote{4+1 Model} by Kruchten~\cite{kruchten1995the}, where an architecture consists of multiple views: A logical view, a development view, a process view, and a physical view.
With this model, we have full freedom to choose representations for these views, like PlantUML or Mermaid diagrams.
Another promising option is the Palladio Component Model~\cite{becker2009palladio}, which would allow us to run quantitative architecture evaluations.
This model specifies the exact format for its different views.
To some degree, we can also convert an architecture from one model into another.
The exact format used will be the subject of our research.

To generate the architectures, we will employ multiple techniques and technologies:
For tasks that require a deep understanding of the requirements, like cutting the architecture into components, we will employ LLMs.
Where feasible, we will also consider other modern AI technologies.
Some tasks are very technical and based on well-structured data, like the process view, which explains the execution path the use cases take through the components.
For such tasks, we will instead apply classic algorithms, which do not have the engineering difficulties that modern AI technology has, like shaky and unpredictable results.

Furthermore, we want to extract the architectural design decisions taken during the generation and model them as architecture design decision records (ADRs)~\cite{vanHeesch2012ADR,kopp2018MADR}.
They will give a quick overview of the essence of each architecture candidate.
They will give hints to each architecture candidate on potential refinements that can be made.
Moreover, they will document the architect's decisions during this process instead of implicit or unconscious decisions taken by the tooling.

\subsection{Architecture Candidate Evaluation}\label{sub:candidate-eval}
After the generation, we want to assist the architect in selecting the best-fitting architecture candidate for the system he designs.
We aim to evaluate the architecture candidates and find their pros and cons.
We want to automate established architecture evaluation technologies like ATAM~\cite{kazman2000atam} to achieve this.
In addition, we will calculate quality metrics on the generated architecture candidates.
By conducting a quantitative analysis, we aim to find potential performance hotspots.
The selection of the architecture evaluation methods will be part of our research:
We will consider multiple evaluation techniques and assess the ability to conduct them automatically.
Furthermore, we will compare the aspects of the architecture they evaluate and select the techniques in a manner that makes sure we evaluate a wide range of aspects to give a comprehensive architecture evaluation overall.
Whether an accurate and comprehensive automatic evaluation is possible will be an outcome of our research.
If this proves to be difficult, we will fall back to a semi-automated evaluation.

\subsection{Architecture Candidate Selection}
Finally, we present the generated architecture candidates, design decisions, and evaluations to the architect.
With this information given, the architect can understand the advantages and disadvantages of each architecture candidate.
The architect can factor in potential implicit requirements that are not documented.
Such requirements might exist for various reasons:
Maybe no one found it necessary to document them because all the people involved were aware of them.
Another possibility could be that those are not strictly considered requirements, but the architect already suspects they might become such.

At this stage, the architect might find that none of the generated architecture candidates quite fit their needs.
In this case, the architect can refine the architecture candidates similarly as they refined the scenarios and the domain model:
Using prompts, the architect can tell the tooling what aspects of the architecture candidate need to be changed.
They might even ask for the generation of further architecture candidates that they can assess and refine as well.
With all these choices available and with the evaluation assistance, the architect can select one of these candidates as the final architecture. 

\subsection{Further Iterations}
After the architecture is selected, the development team can continue implementing it.
We assume that, in practice, at some point, the requirements will change.
This might either be planned, e.g., when the next iteration of an agile process begins, or unplanned, e.g., due to a misconception in the requirements.
In this case, we want to support the iteration of the proposed process:
This means that, as an additional requirement in the iteration, the generated architecture candidates should differ as little as possible from the current software architecture.
Each change in the architecture might imply expensive changes in the source code, and we want to reduce them as much as possible.
But sometimes, architectural changes are necessary to keep up with the project's quality requirements.
This means that we cannot restrict the process to retain the current architecture at all costs.
Instead, the tooling needs to weigh the quality implications against the expenses for an architectural change.
These decisions will -- to some degree -- be presented to the architects, with these decisions factored into the presented data.
This will also enable this process for agile environments, which are very common in the industry.

\section{Exploratory Analysis}\label{sec:exploratory-analysis}
We did an exploratory analysis with chat versions of LLaMA2 70-B and GPT-3.5 to generate domain models of text requirements.
For this, we supplied 91 requirements from the MobSTr-dataset~\cite{mobstr_2021_06} and asked the model to generate a PlantUML domain model.
The exact procedure can be found in our GitHub\footnote{\url{https://github.com/qw3ry/requirements2architecture/tree/exploration}}.
Both models could identify concepts from the requirements.
It appears that the prompts were misunderstood, though -- instead of modeling the domain, the LLMs modeled the system itself.
Still, both LLMs showed an understanding of the requirements.
While both models could generate valid PlantUML, LLaMA did not create relations between the concepts, even though it claimed it did.
GPT handled this task well, though.
In a small iteration, we could improve the results of both models substantially with basic prompt engineering.

With these tests, using general, unmodified LLMs, we can assume that LLMs are generally capable of understanding application domains.
We must invest some work into adjusting these models, prompt engineering, and translating the LLM output into the chosen format.

\section{Planned Evaluation}\label{sec:evaluation} %
We plan to evaluate the process as a whole with two studies, and we aim to evaluate both the technical quality of the outcome and the impact on the development process.
In addition, we plan to evaluate the intermediate steps of the process thoroughly, especially regarding the technical quality of the intermediate artifacts.
We hope to apply these intermediate studies' findings to improve the final evaluation process.

\subsection{Evaluation Based on Reference Architectures}
We want to take well-established reference architectures like T2-Project~\cite{Speth2022T2}, TeaStore~\cite{vonkistowski2018teastore}, and SockShop, re-engineer their requirements, and apply our process.
We will then manually apply architecture evaluation techniques to compare the reference architecture to the generated architecture candidates.
Because we are not constrained to automatic evaluation techniques, we can evaluate the architectures more in-depth when compared to the automatic evaluation described in \cref{sub:candidate-eval}.
In addition, we will examine the generated architecture candidates and compare them to the reference architecture to detect potential flaws that the more formal architecture evaluation techniques missed.
Furthermore, we will ask experts to create architectures for the evaluated requirements and then rate the generated architecture candidates for their fit to the requirements.
With the architecture created by the expert, we have a baseline of what an architect in a typical project is roughly capable of.

The used LLM might have been trained with the reference architectures previously.
Thus, we will carefully check whether the LLMs only reproduce the reference architectures, or incorporate new ideas.
If we find that this limitation applies, we will apply the described evaluation instead with a new set of requirements.

\subsection{Industrial Field Study}
We want to conduct an industrial field study to evaluate our process in a real-world environment.
For this evaluation, we will use the requirements of a project and apply our process.
We will then interview the project members on the quality of the generated architecture and the quality of the actual architecture of the project.

Secondly, we will deploy our process for some time in the pro\-ject's ongoing development.
Afterward, we will interview the pro\-ject members again, this time focusing on the challenges of the process implementation and the advantages and disadvantages of the process.
This evaluation will be very open, and we aim to find challenges in usability, practicality, and the overall hurdles of implementing this process.
We want to evaluate the effectiveness and efficiency of the process.

Lastly, we want to apply the Technology Acceptance Model (TAM~\cite{davis1985technology}) to evaluate the attitude towards and the intentions to use this process.

\section{Conclusion}\label{sec:conclusion}
We propose research for a new process that aids architects in their daily work.
This process spans the whole creation of a new architecture and can support the adaption of an existing architecture to new requirements.
We believe this process can be the basis for a considerable step forward in the quality of industrial software architectures.

\bibliographystyle{ACM-Reference-Format}
\bibliography{bibliography}

\end{document}